\documentclass[11pt]{article}
\usepackage{fullpage}
\usepackage{graphicx}           
\usepackage{bm}                 
\usepackage{amsmath}            
\usepackage{amsfonts}
\usepackage{amssymb}
\usepackage{latexsym}           

\usepackage{natbib}
\bibpunct[,]{[}{]}{,}{n}{,}{,}

\newcommand{\ket}[1]{\mbox{$|#1\rangle$}}

\newcommand{\identity}{\leavevmode\hbox{\small1\kern-3.2pt\normalsize1}}

\begin{document}

\title{Coined quantum walks on percolation graphs}
\author{Godfrey Leung\thanks{Current address: School of Physics and Astronomy, University of Nottingham, University Park, Nottingham, NG7 2RD, United Kingdom. Contact: \texttt{ppxgl@nottingham.ac.uk}},
	Paul Knott,
	Joe Bailey\thanks{Current address: Centre for Mathematics and Physics in the Life Sciences and Experimental Biology (CoMPLEX), University College London, Gower Street London WC1E 6BT, United Kingdom.  Contact: \texttt{joe.bailey.09@ucl.ac.uk}},
    and Viv Kendon\thanks{Contact: \texttt{V.Kendon@leeds.ac.uk}}\\
\textit{School of Physics and Astronomy, University of 
             Leeds, LS2 9JT, United Kingdom.}}

\date{\today}

\maketitle

\begin{abstract}
Quantum walks, both discrete (coined) and continuous time,
form the basis of several quantum algorithms and have been
used to model processes such as transport in spin chains and
quantum chemistry.
The enhanced spreading and mixing properties of quantum walks
compared with their classical counterparts have been well-studied
on regular structures and also shown to be sensitive to defects and
imperfections in the lattice.
As a simple example of a disordered system, we consider percolation
lattices, in which edges or sites are randomly missing, interrupting
the progress of the quantum walk.
We use numerical simulation to study the properties of 
coined quantum walks on these percolation lattices in one and two
dimensions.
In one dimension (the line) we introduce a simple notion of
quantum tunneling and determine how this affects the properties of
the quantum walk as it spreads.
On two-dimensional percolation lattices, we show how the spreading
rate varies from linear in the number of steps down to zero,
as the percolation probability decreases towards the critical point.
This provides an example of fractional scaling in quantum walk dynamics.
\end{abstract}


\tableofcontents


\section{Introduction}
\label{sec:intro}

Quantum versions of random walks have been extensively studied since
they were introduced in the context of quantum algorithms
\cite{farhi98a,aharonov00a,ambainis01a}.
A classical random walk is essentially a diffusion process, in 
which spreading from the initial configuration occurs as a series
of random steps. 
In a quantum walk, quantum coherences replace the diffusion, which
can lead to both faster and slower spreading from the starting point of
the quantum walk.
Faster spreading provides the speed up in many quantum walk algorithms.
Exponential speed up has been proved for certain transport problems:
\citeauthor{kempe02a}~\cite{kempe02a,kempe02c} showed how a quantum
walk can cross a hypercube much more efficiently than a classical 
random walk, while
\citeauthor{childs02a}~\cite{childs02a} have produced a scheme for a
continuous time quantum walk that can find its way across a particular
``glued trees'' graph exponentially faster than any classical algorithm.
On the other hand, the phenomenon of Anderson localization of quantum
particles is well-known, and has been related to continuous-time quantum walks
by \citeauthor{keating06a}~\cite{keating06a}.
Studies by \citeauthor{krovi05a}~\cite{krovi05a,krovi06a,krovi07a}
extend this to coined quantum walks, and highlight the importance of
symmetry in determining the properties of quantum walks.
\citeauthor{stefanik07a} \cite{stefanik07a,stefanik08a}
provide a way to characterize the localization of unbiased quantum walks on
regular lattices, showing how it depends on the topology of the lattice,
the coin operator and chosen initial state.

Localization can also be exploited, as in the quantum walk search algorithm,
where a marked state perturbs the uniform distribution causing the quantum
walk to converge onto it.  \citeauthor{shenvi02a} \cite{shenvi02a}
proved that basic quantum walk searching is quadratically faster,
the maximum speed up that can be achieved for unordered searching
\cite{bennett97a}.  This quadratic improvement is only achievable for
a well-connected database though, on a two-dimensional grid it is
still open whether the full quadratic speed up is possible.
For a recent review of quantum walk searching and related open
problems, see \citeauthor{santha08a}~\cite{santha08a}.

Quantum walks also provide useful models of physical phenomena
such as quantum state transfer in spin chains \cite{bose03a}, or
energy transport in biomolecules \cite{mohseni08a}.
Here the goal is to transfer a quantum state or an excitation,
with high fidelity being more important than high speed.
Certain combinations of parameters can achieve perfect state
transfer (reviewed in \cite{kendon10b}), while a small amount of 
noise can produce robust near-perfect transfer in more disordered
systems \cite{mohseni08a}.
There is no general formula for predicting when a quantum walk will
be more efficient than its classical counterpart, and the 
wide-ranging studies to date (e.g., \cite{agliari09a})
only scratch the surface of the possible configurations and
parameters available.

In this work, we are motivated by using quantum walks as models of
transport phenomena, so we are interested in how fast they spread
from a localized starting state.
In order to have a simple, but flexible,
concrete setting for our work, we concentrate on the behaviour
of quantum walks on percolation lattices, i.e., Cartesian lattices
with vertices or edges randomly missing, that thus limit the paths the
quantum walker can take.
Edge (bond) and site (vertex) percolation lattices
are defined by the probability $p$ of an edge or site respectively
being present in the structure.  In two or more dimensions, as the proportion
of missing edges or sites decreases, around some critical probability $p_c$
the lattice changes from a collection of disconnected sections to
almost all connected.  Moreover, there will in general be a path
from one side of the lattice to the opposite side for $p>p_c$,
an obvious qualitative change in the transport properties of the lattice.
The two types of lattice (with either site or bond percolation) have similar
properties, but different critical parameters.
They are used to model the behaviour of phenomena
as diverse as the spread of contagious diseases,
the propagation of forest fires, and the movement of oil deposits
in porous rocks.
They have been studied in great detail and, where analytical solutions
are not available, efficient numerical methods have been developed
\cite{djordjevic82a,gebele84a}.
For an accessible introduction to the properties and uses of
percolation lattices, see \citeauthor{stauffer1994}~\cite{stauffer1994}.
For this work, we are simply using the percolation lattices as a 
substrate for the quantum walks.  One of the most basic questions we
can ask is how the quantum walk behaves as
the density of edges or sites is varied.
In related work,
\citeauthor{xu08a} \cite{xu08a} study quantum walks on
{E}rd\"{o}s-{R}\'{e}nyi networks, which also have randomly placed edges
similar to percolation lattices, but without the underlying regular
structure of a lattice.  The limiting case is thus a walk on the complete
graph, rather than the more familiar walk on a line or grid.

Like classical random walks, quantum walks come in both discrete time
\cite{aharonov92a,watrous98a,aharonov00a,ambainis01a},
and continuous time \cite{farhi98a} versions.  
Here we investigate only discrete time (coined) quantum walk dynamics,
though we expect the continuous time dynamics to be qualitatively similar in
many respects.  
We first review the properties of a discrete quantum walk on 
the line.
For the first half of the paper, we work mainly with edge (bond)
percolation lattices.
In one spatial dimension (the line), just a single missing edge
is enough to halt the progress of both quantum walks and random walks.
So we explore
two less restrictive variations on this theme: dynamic gaps, where
the missing edges change with each time step \cite{romanelli04a};
and tunneling, a simple
model of the quantum phenomenon whereby a quantum particle
is able to pass through a narrow barrier with some small probability.
There is also no (non-trivial) phase transition in one dimension
for the percolation lattice, effectively $p_c=1$.

We follow these one-dimensional examples with studies
on two dimensional lattices.
In general there is now more than one possible path between two points,
allowing for non-trivial behaviour by the quantum walker,
even with gaps fixed throughout the walk.
Bond (edge) percolation in a two-dimensional Cartesian lattice has a
phase transition at $p_c=0.5$.
We also compare site percolation, where $p_c=0.5927\dots$ for two dimensions
\cite{djordjevic82a,gebele84a}.
We then discuss our results in the final section.

\section{Walk on the line}
\label{sec:line}

A discrete quantum walk on a line can be defined in direct analogy with
a classical random walk, where the walker repeatedly steps one unit forward
or back based on the outcome of a random coin toss.
The quantum walker has a quantum coin that will in 
general be in a superposition of ``forward'' and ``back'',
and the quantum walker can step into a superposition of positions on the line,
based on the state of the coin.
However, the coin toss is not random, since pure quantum dynamics need
to be unitary, but if measured, the outcome of the measurement
would be random, like the classical coin.

To fix our notation, we define a two-dimensional quantum coin with basis states 
$\ket{c}$ with $c \in \{\pm 1\}$,
and label the position basis states $\ket{x}$
by the integer location on the line $x$. 
A general state of the quantum walker can thus be written
\begin{equation}
\ket{\psi(x,t)} = \sum_{c,x} a_{c,x}(t)\ket{c,x},
\end{equation}
where $a_{c,x}(t)$ is the complex amplitude associated with the walker being
at position $x$ with the coin in state $c$ at time step $t$,
satisfying the normalization condition $\sum_{c,x} |a_{c,x}(t)|^2 = 1$.
Here we have written $\ket{c,x}\equiv\ket{c}\otimes\ket{x}$ for
combined basis states.
The evolution of the walk is governed by a coin
operator $\mathbf{C}_2$ that acts on the quantum coin at each step of the walk.
The simplest and most commonly used coin operator is the Hadamard operator,
\begin{equation}
\mathbf{C}_2^{(\text{Had})}=\frac{1}{\sqrt{2}}\left( \begin{array}{cc}
        1 & 1\\
        1 & -1
        \end{array} \right).
\label{eq:had}
\end{equation}
After ``tossing'' the coin with the coin operator, the particle moves to 
adjacent positions according the the coin state; this is expressed 
mathematically as a conditional shift operator
\begin{eqnarray}
&&\mathbf{S}\ket{{-1},x} = \ket{{-1},x-1}\nonumber\\
&&\mathbf{S}\ket{{+1},x} = \ket{{+1},x+1} .
\label{eq:cshift}
\end{eqnarray}
One step of the quantum walk is produced by the unitary operator
$\mathbf{U} = \mathbf{S}(\mathbf{C}\otimes\identity_x)$.
A quantum walk of $t$ steps starting from an initial state
$\ket{\psi_0}$ is thus 
\begin{equation}
\ket{\psi(x,t)} = U^t\ket{\psi_0}.
\label{eq:psit}
\end{equation}

\begin{figure}[tbh!]
  \begin{minipage}{1.0\columnwidth}
    \begin{center}
        \resizebox{0.55\columnwidth}{!}{\rotatebox{0}{\includegraphics{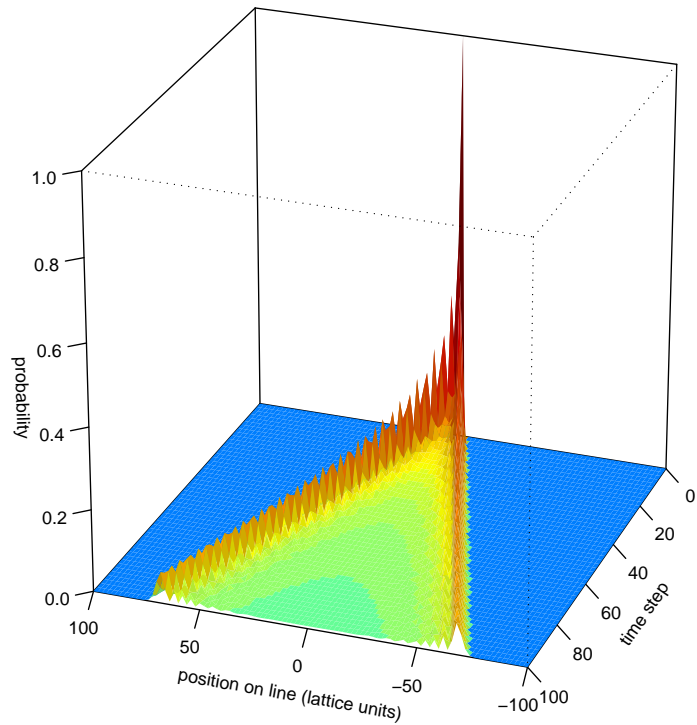}}}
    \end{center}
  \end{minipage}
  \caption{Probability distribution for a quantum walk on the line
	over 100 steps showing the time evolution, using a Hadamard coin,
	equation (\ref{eq:had}),
        and a symmetric initial state $(\ket{{+1},0}+i\ket{{-1},0})/\sqrt{2}$.
	Only positions with non-zero probability of occupation are shown,
	since odd positions are unoccupied at even time steps and vice versa.}
  \label{fig:1dwalk_evol_normal}
\end{figure}

The position probability distribution of a quantum walk on a line,
originally solved analytically by
\citeauthor{ambainis01a} \cite{ambainis01a,nayak00a},
is by now well-known, the time evolution over 100 steps with a
Hadamard coin operator and initial state of 
$\frac{1}{\sqrt{2}}(\ket{{+1,0}}+i\ket{{-1,0}})$ 
is shown in figure \ref{fig:1dwalk_evol_normal}.
The double-peaked spreading is quite different to the binomial distribution
of a classical random walk, and expands at a linear rate, giving a
quadratic speed up over the $\sqrt{t}$ spread of the binomial distribution.

Since the walk on the line is amenable to analytic treatment, it
has been exhaustively studied by many authors over over the past ten years.
Both path-counting and  Fourier transform methods were given by
\citeauthor{ambainis01a} \cite{ambainis01a}, and path counting (path integrals)
were further refined by \citeauthor{carteret03a} \cite{carteret03a}.
A method using the algebra of the matrix operators was presented by
\citeauthor{konno02a} \cite{konno02a,konno02c}
and the tools of classical optics were
adapted to quantum walks by \citeauthor{knight03c} \cite{knight03c}.
\citeauthor{romanelli03a} \cite{romanelli03a}
separate the dynamics into Markovian and
interference terms, which they relate to the dynamics of a kicked rotor.
The coin operator may be generalised to any SU(2) operator,
\begin{equation}
\mathbf{C}_2^{(\text{gen})}=\left( \begin{array}{cc}
        \sqrt{\eta}& e^{i\theta}\sqrt{1-\eta}\\
        e^{i\phi}\sqrt{1-\eta}& -e^{i(\theta+\phi)}\sqrt{\eta}
        \end{array} \right),
\label{eq:genH}
\end{equation}
where $0\le\theta,\phi\le\pi$ are arbitrary phase angles and
the coin bias is $0\le\eta\le 1$.
If we set $\eta=0.5$  and $\theta=\phi=0$, we obtain
the Hadamard coin operator of equation (\ref{eq:had}).
The initial state of the coin can be similarly varied,
\begin{equation}
\ket{\psi(0)} = \sqrt{\eta'}\ket{{+1,0}}+e^{i\theta'}\sqrt{1-\eta'}\ket{{-1,0}},
\end{equation}
where $0\le\theta'\le\pi$ is an arbitrary phase angle and
the coin initial state bias $0\le\eta'\le 1$.
As was shown by \citeauthor{bach02a} \cite{bach02a}, it is sufficient to 
specify one phase to obtain the full range of behaviour from
the subsequent walk on the line starting from the origin.
It is convenient for our purposes here to place the phase factor in
the coin initial state ($\theta'$), and deal only with
coin operators of the form
\begin{equation}
\mathbf{C}_2^{(\eta)}=\left( \begin{array}{cc}
        \sqrt{\eta} & \sqrt{1-\eta}\\
        \sqrt{1-\eta} & -\sqrt{\eta}
        \end{array} \right).
\label{eq:bias}
\end{equation}
Varying the phase alone ($\eta=\eta'=0.5$) produces biased
walk distributions with a greater probability of the walker being
found in the positive (negative) directions for $\theta' = 0(\pi)$.
Setting $\theta'=\pi/2$ gives the unbiased walk shown in
figure \ref{fig:1dwalk_evol_normal}.
Varying the bias $\eta'$ in the initial state also controls how skew or
symmetric the resulting walk turns out to be 
\cite{tregenna03a,konno04a}.   Varying the bias in the coin operator
$\mathbf{C}_2^{(\eta)}$ controls how fast the walk spreads.  The
extreme cases are $\eta=1$, for which the walker hops along the line
without any reversing, in one or both directions
as indicated by the initial coin state, and $\eta=0$, for which the walk
oscillates back and forth between the initial state and nearest
neighbouring positions, making no further progress along the line at all.
There have also been extensive studies of decoherence applied to quantum
walks, both analytic and numerical, see review by
\citeauthor{kendon06b} \cite{kendon06b}, and references therein.

Numerical simulation of a quantum walk is a straightforward
evaluation of equation (\ref{eq:psit}) for chosen parameters.
In high level computational environments such as Matlab, it takes only
a few lines of code.  However,
we wrote our own routines in C and C++ to ensure they were efficient enough to
to run up to at least 10,000 time steps for the walk on the line, and
over 100 steps for the walk on a 2D lattice (the memory required scales
linearly in the number of lattice sites).
It was also necessary to repeat the simulations over a sufficient number
of random percolation lattices to ensure reasonable statistics were 
obtained.  The largest of our simulations took over a week on a 3GHz
processor to complete this averaging for 5,000 random lattices.
This was in fact largely due to the time required to generate the random
numbers for specifying the percolation lattice (one per walk).
We used routines from \citeauthor{park98a} \cite{park98a} freely available
online for generating the random numbers, although the quality of the
random numbers is not particularly important for this type of simulation.

\subsection{Dynamic gaps}

\begin{figure}[tbh!]
  \begin{minipage}{1.0\columnwidth}
    \begin{center}
        \resizebox{0.65\columnwidth}{!}{\rotatebox{0}{\includegraphics{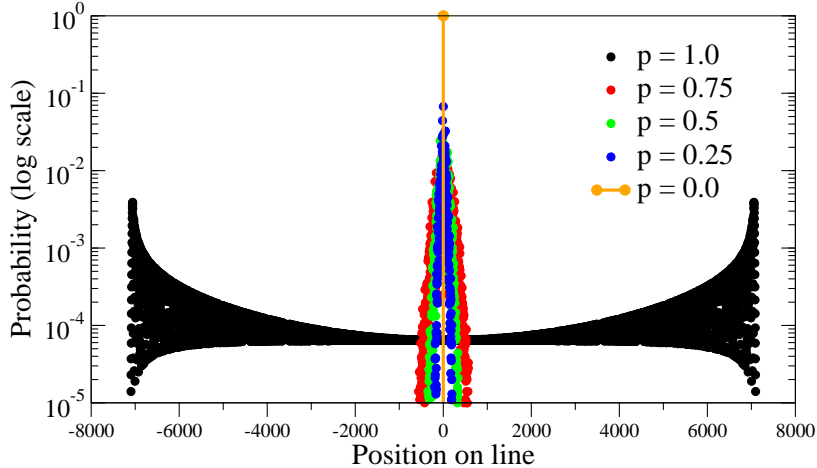}}}
    \end{center}
  \end{minipage}
  \caption{Probability distribution after 10,000 time steps
	for a quantum walk on the line with dynamic gaps
	where the probability of an edge being present is
	$p=1$ (black), 0.75 (red), 0.5 (green), 0.25 (blue) and 0.0 (orange),
        using a Hadamard coin, equation (\ref{eq:had}),
        and a symmetric initial state $(\ket{{+1},0}+i\ket{{-1},0})/\sqrt{2}$.
        Only even positions shown since odd positions are unoccupied.}
  \label{fig:1dwalkprobspread_dynamic}
\end{figure}

If edges are missing from the line, neither the classical nor
the quantum walkers have any way to move beyond the gaps,
and are thus constrained to the section of the line on which they start.
A more interesting problem is the effect
of gaps that change location after each time step.  The gaps are then only a
temporary barrier to the walkers' progress.  In order to apply a quantum 
walk to a dynamically changing structure, the unitary evolution 
becomes dependent on both position and time.  The simplest way to
accomplish this is to modify the shift operator $\mathbf{S}$, equation
(\ref{eq:cshift}) while keeping the coin operation as before,
equation (\ref{eq:had}).
The variable shift operator $\mathbf{S}(x,t)$ is then 
composed of the following three variants in addition to the regular
shift in equation (\ref{eq:cshift}) that is applied at positions with
both edges present.  If the edge in the plus direction is missing,
\begin{eqnarray}
&&\mathbf{S_+}\ket{{-1},x} = \ket{{-1},x-1}\nonumber\\
&&\mathbf{S_+}\ket{{+1},x} = \ket{{-1},x} ,
\label{eq:cshift+}
\end{eqnarray}
if the edge in the minus direction is missing,
\begin{eqnarray}
&&\mathbf{S_-}\ket{{-1},x} = \ket{{+1},x}\nonumber\\
&&\mathbf{S_-}\ket{{+1},x} = \ket{{+1},x+1} ,
\label{eq:cshift-}
\end{eqnarray}
and if both edges are missing,
\begin{eqnarray}
&&\mathbf{S_0}\ket{{-1},x} = \ket{{+1},x}\nonumber\\
&&\mathbf{S_0}\ket{{+1},x} = \ket{{-1},x} .
\label{eq:cshift0}
\end{eqnarray}
Since $\mathbf{S_+}$ and $\mathbf{S_-}$ occur in pairs either side
of gaps, the overall $\mathbf{S}(x,t)$ is unitary.
This model was first studied 
by \citeauthor{romanelli04a} \cite{romanelli04a}, who found that, as the
proportion of gaps increases ($p$ decreases), 
the distribution of the quantum walk tends to the classical binomial spread,
after a characteristic time that scales as $1/(1-p)$.
This is illustrated in figure \ref{fig:1dwalkprobspread_dynamic},
(which has been plotted for single runs, i.e., no averaging over
the different possible sequences of missing gaps).
Dynamic gaps thus act as a type of decoherence that transforms the quantum
walk behaviour into classical random walk behaviour.
This is thus an example of behaviour with a single scaling of
$\sqrt{t}$, but a prefactor that varies depending on the 
parameters (the percolation probability in this case).  
\citeauthor{romanelli04a} \cite{romanelli04a} compare this with
decoherence due to periodic and random measurements (non-unitary
evolution) applied to both coin and position, which also
reduces the scaling to $\sqrt{t}$ but with larger prefactors.
Similar behaviour was also found by \citeauthor{brun02c} \cite{brun02c,brun02a}
for decoherence applied only to the coin driving a quantum walk on the line.
\citeauthor{brun02c} solved their model analytically for the spreading rate.
They found that the limiting distribution scaled classically
as $\sqrt{t}$, but the prefactor was larger and determined by
the decoherence rate.

\subsection{Quantum Tunneling}
\label{ssec:qtun1}

Tunneling is a generic property of quantum systems.  One of the easiest
demonstrations uses light undergoing total internal reflection in
a block of glass. If another block of glass is brought close to the
surface where the reflection is taking place, some of the light
can be seen to be transmitted even when there is a small air gap
between the blocks.  The probability of finding photons outside
the glass at the point of reflection isn't zero, it
falls away exponentially from the interface (the evanescent wave).
If a second glass block
is placed close to the first one, the photons return to propagating
in a normal wave-like fashion in the second block.
The amplitude of the transmitted wave is given by the amplitude
of the evanescent wave at the second interface.

We can make a simple model of quantum tunneling by allowing
the quantum walker to hop over a broken link using the biased
coin $\mathbf{C}_2^{(\eta)}$ from equation (\ref{eq:bias}).
Changing $\eta$ is equivalent to 
changing the size of the air gap between the glass blocks.
The quantum walk dynamics will now have a
position dependent coin operator, while the shift operator
remains as in equation (\ref{eq:cshift}).
We apply $\mathbf{C}_2^{(\eta)}$ at the sites on both
sides of the missing edge to create the tunnel.
Using $0 < \eta < 0.5$ corresponds to tunneling with a reduced
probability of transmission, while $0.5 < \eta \le 1$ would give us 
enhanced transmission, which we don't have a use for in this model.
A coin operator composed of a combination of these two operators,
$\mathbf{C}_2^{(\text{Had})}$ and $\mathbf{C}_2^{(\eta)}$ 
is easily seen to be unitary, because the full coin operator is a block
diagonal matrix, $\mathbf{C}\otimes\identity_x$, and the position
dependence simply inserts the different $\mathbf{C}$ into the 
appropriate blocks.

\begin{figure}[tbh!]
  \begin{minipage}{1.0\columnwidth}
    \begin{center}
        \resizebox{0.55\columnwidth}{!}{\rotatebox{0}{\includegraphics{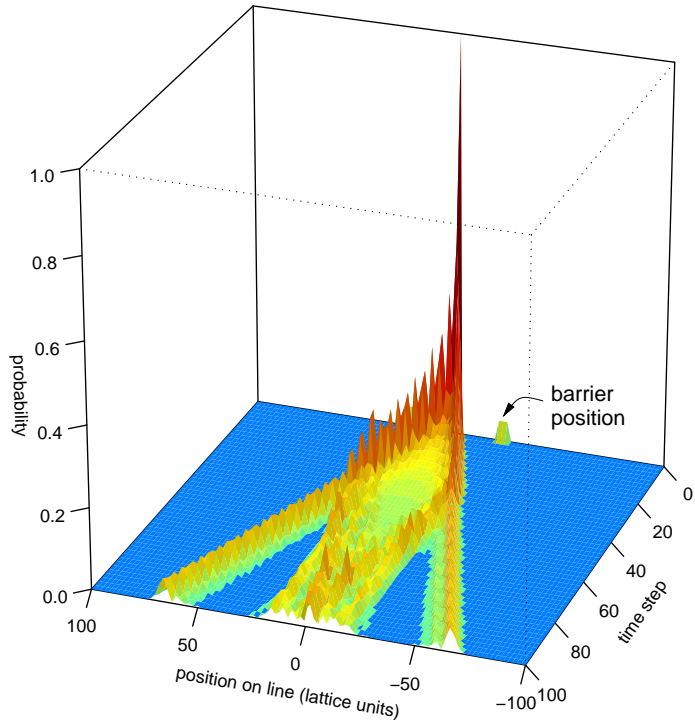}}}
    \end{center}
  \end{minipage}
  \caption{Probability distribution for a 1D walker on a line
	showing the time evolution over 100 steps
	using a Hadamard coin, equation (\ref{eq:had}),
	and a symmetric initial state $(\ket{{+1},0}+i\ket{{-1},0})/\sqrt{2}$.
	The tunneling strength $\eta = 0.25$, and
	gaps five edges wide are placed at $\pm$20, as indicated by
	the labeled mark at $t=0$.
	Only positions with non-zero probability of occupation are shown,
	since odd positions are unoccupied at even time steps and vice versa.}
  \label{fig:1dwalk_T_barrier}
\end{figure}

To see how the tunnelling works in more detail,
suppose there is a gap between position $x$ and position $x+1$.
If the walker arrives at $x$ from $x-1$, the coin will necessarily
be in state $\ket{{+1}}$.  Applying a step of the walk, 
$\mathbf{U(\eta)} = \mathbf{S}\mathbf{C}_2^{(\eta)}$ to the state $\ket{{+1,x}}$ gives
\begin{equation}
\mathbf{S}\mathbf{C}_2^{(\eta)}\ket{{+1,x}} = 
\sqrt{1-\eta}\ket{{-1,x-1}} + \sqrt{\eta}\ket{{+1,x+1}}.
\end{equation}
A fraction $\sqrt{\eta}$ of the amplitude has thus crossed the gap.
We now need to apply the same operator $\mathbf{U(\eta)}$ to this
amplitude at position $x+1$,
\begin{equation}
\mathbf{S}\mathbf{C}_2^{(\eta)}\sqrt{\eta}\ket{{+1,x+1}} = 
\sqrt{\eta(1-\eta)}\ket{{-1,x}} + \eta\ket{{+1,x+2}}.
\end{equation}
Thus after two steps, only a fraction $\eta$ of the amplitude has
progressed completely past the gap.  This compares with
the expected fraction of half the amplitude that would have
reached this position after two steps of the walk if the gap weren't there.
Rigorously defining the probability of tunnelling through the
gap requires a more thorough treatment of the delocalized
quantum walker, for example, using a
traveling wave incident on the gap, rather than localised 
position states.   The coin operators then act like beam splitters,
see \citeauthor{knight03c} \cite{knight03c}.
For our purposes we don't need a precise definition, the biased coin gives us
a simple way to interpolate smoothly between an impassable gap and
the unobstructed walk, which allows us to explore the intermediate
behaviour qualitatively.
By placing a pair of gaps at $\pm 20$ and starting the quantum walk at
the origin, the reflection and transmission at the gaps
is nicely illustrated in figure \ref{fig:1dwalk_T_barrier}.
A different model of tunneling to account for imperfect shift 
operations in a quantum walk on an unbroken line has been studied
numerically by \citeauthor{dur02a} \cite{dur02a}, and analytically
by \citeauthor{annabestani10a} \cite{annabestani10a}, who
concluded that in this scenario, the extra shifting does
not destroy the quantum spreading rate.

\begin{figure}[tbh!]
  \begin{minipage}{1.0\columnwidth}
    \begin{center}
        \resizebox{0.65\columnwidth}{!}{\rotatebox{0}{\includegraphics{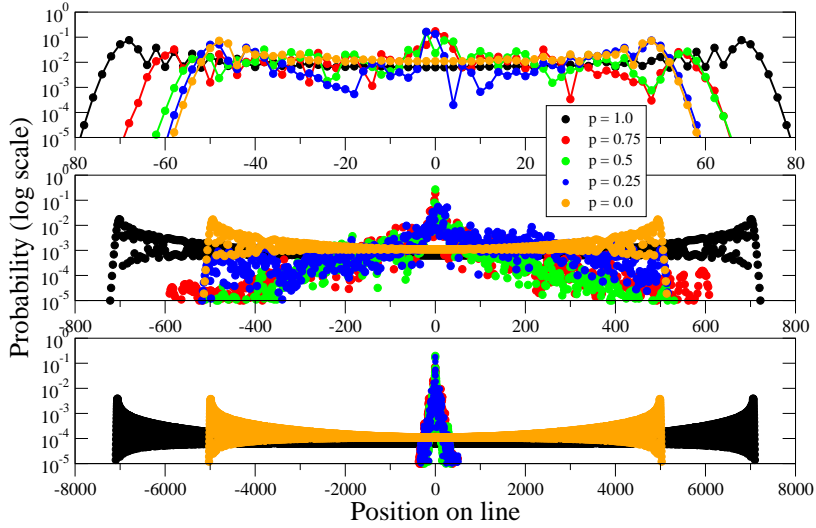}}}
    \end{center}
  \end{minipage}
  \caption{Probability distribution for a 1D walker on a line with
	static gaps and a tunneling strength of $\eta=0.25$
	after 100 steps (top), 1,000 steps (middle), and 10,000 steps (bottom)
	using a Hadamard coin, equation (\ref{eq:had}),
	and a symmetric initial state $(\ket{{+1},0}+i\ket{{-1},0})/\sqrt{2}$,
	where the probability of an edge being present is
	$p=1$ (black), 0.75 (red), 0.5 (green), 0.25 (blue) and 0.0 (orange).
	Only even positions shown since odd positions are unoccupied.}
  \label{fig:1dwalk_T_static}
\end{figure}

Figure \ref{fig:1dwalk_T_static} shows the effect of tunneling on
a quantum walk on a line with randomly placed static gaps and a tunneling
strength of $\eta=0.25$ (half the normal unbiased coin transmission rate).
The number of missing edges is controlled by the probability $p$ of any
given edge being present.
After 100 steps, the spreading is essentially still at the fully
quantum rate for all values of $p$.
By 1000 steps a small central peak has appeared at the expense of
the leading edges, while for 10,000 steps the intermediate values of $p$
have reduced the walk to classical-like distributions (like classical
random walks on the unbroken line).
The random distribution of gaps in the percolation lattice removes
a random subset of the possible same-length paths between any two points.
This slowly destroys the long-term coherence required for the full
quantum spreading rate.
After 10,000 steps, only $p=1$ (no gaps) and $p=0$ (all tunneling) show 
double-peaked quantum speed up.
\begin{figure}[tb!]
  \begin{minipage}{1.0\columnwidth}
    \begin{center}
        \resizebox{0.65\columnwidth}{!}{\rotatebox{0}{\includegraphics{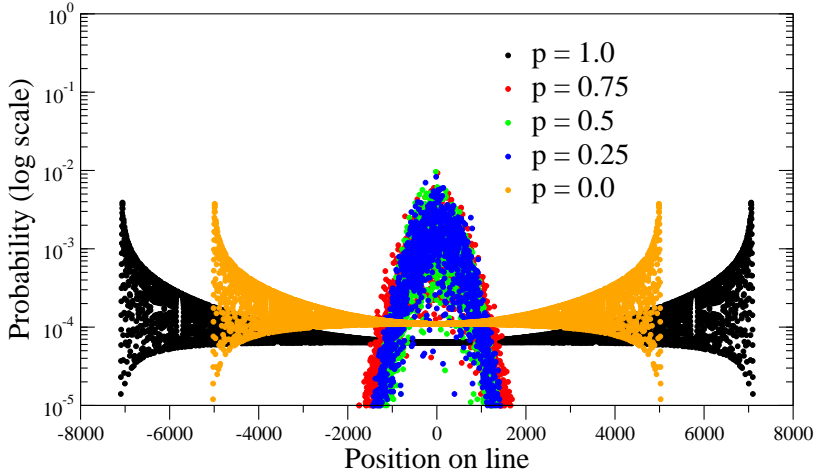}}}
    \end{center}
  \end{minipage}
  \caption{Probability distribution for a 1D walker on a line with
	dynamic gaps where the probability of an edge being present is
	$p=1$ (black), 0.75 (red), 0.5 (green), 0.25 (blue) and 0.0 (orange),
	and a tunneling strength of $\eta=0.25$ 
	after 10,000 steps using a Hadamard coin, equation (\ref{eq:had}),
	and a symmetric initial state $(\ket{{+1},0}+i\ket{{-1},0})/\sqrt{2}$.
	Only even positions shown since odd positions are unoccupied.}
  \label{fig:1dwalk_T_dynamic}
\end{figure}
Figure \ref{fig:1dwalk_T_dynamic} shows the combination of the same
dynamic gaps as in figure \ref{fig:1dwalkprobspread_dynamic} and
tunneling with $\eta=0.25$.  The tunneling clearly enhances the 
progress of the walk considerably, but the dynamic gaps still cause decoherence
and a consequent return to a classical (binomial) shaped distribution for the
intermediate values of $p$.  However, the width of the distribution is
significantly larger than the classical random walk distribution on
the full line.
The standard deviation for the classical random walk
for 10,000 steps is 100, while the standard deviation for the quantum walk
with tunnelling strength $p=0.25$ is 479, for $p=0.5$ it is 397,
and for $p=0.75$ it is 414.  Tunnelling is thus allowing the quantum
coherences to survive for longer.

\begin{figure}[tbh!]
  \begin{minipage}{1.0\columnwidth}
    \begin{center}
        \resizebox{0.55\columnwidth}{!}{\rotatebox{0}{\includegraphics{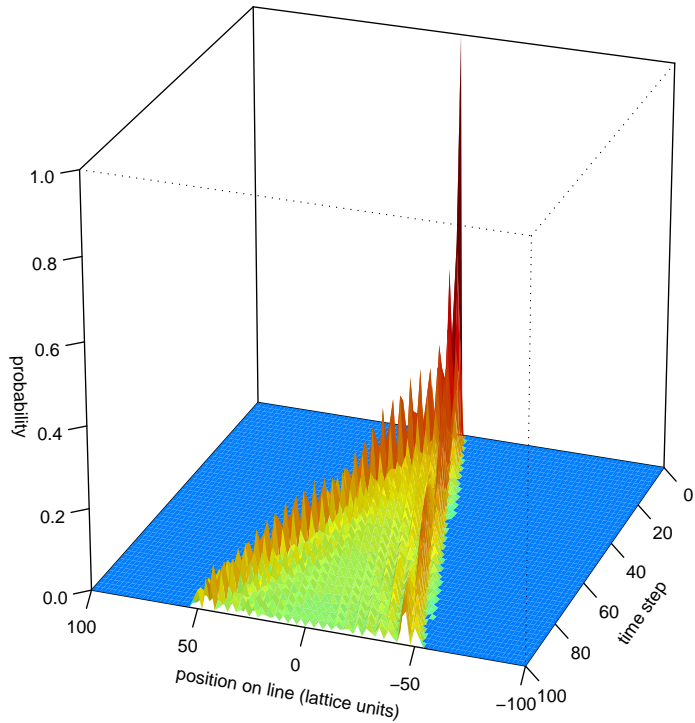}}}
    \end{center}
  \end{minipage}
  \caption{Probability distribution for a 1D walker on a line showing
	evolution over 100 steps using a Hadamard coin, equation (\ref{eq:had}),
	and a symmetric initial state $(\ket{{+1},0}+i\ket{{-1},0})/\sqrt{2}$.
	The tunneling strength $\eta = 0.25$.
	Gaps are placed at every other link.}
  \label{fig:1dwalk_T_fringes}
\end{figure}

In contrast to the decohering effects of random gaps, 
using a regular pattern of static gaps at every other edge
(like a diffraction grating) slows the progress of the walk,
but does not affect its characteristic
quantum shape, see figure \ref{fig:1dwalk_T_fringes}.
This is an illustration of the sensitivity of a quantum walk to symmetry.
Only half that density of random gaps significantly disrupts the quantum walk,
as shown in figure \ref{fig:1dwalk_T_static}.
Regular patterns of varying coin bias have been studied by
\citeauthor{linden09a} \cite{linden09a}, who find quantum spreading 
in most cases, but with widely varying spreading rates.
Generalisation of this model by \citeauthor{shikano10a} \cite{shikano10a}
suggests that localization vs spreading depends on whether the
variation in the coin is governed by a rational or irrational number.


In order to make quantitative comparisons of the effects of
varying the percolation parameter $p$ and the tunneling
strength $\eta$, we calculated the root mean squared (rms) value
of $x$ for the probability distribution, i.e., $x_{\text{rms}} = 
\sqrt(\bar{x^2})$.  We will use the overbar to denote 
averaging over a single quantum walk.
This gives us a measure of how fast the quantum walk is spreading
from its starting point at the origin $x=0$.  We could equally well have
used $\bar{|x|}$, but $x_{\text{rms}}$ 
(often referred to as the standard deviation when the mean is zero)
is more commonly used for this purpose.
Since we want to know about the average
behaviour, we repeated the quantum walk for 1000 different
random distributions of gaps, but for a walk of only 40 steps, to
ensure we obtained reasonable statistics.
We thus have $\langle x_{\text{rms}} \rangle = 
\langle \sqrt(\bar{x^2}) \rangle$
as our measure of the spreading of a quantum walk on a line with 
random gaps.  We will use $\langle . \rangle$ to denote 
averaging over many different randomly generated percolation samples.
It is also useful to examine how variable the
spreading is for fixed $p$ and $\eta$, so we also calculated 
$\sigma(x_{\text{rms}})$, the standard deviation of $x_{\text{rms}}$
over the set of random percolation samples.

\begin{figure}[tbh!]
\begin{minipage}{\textwidth}
  \begin{minipage}{0.48\columnwidth}
    \begin{center}
        \resizebox{0.75\columnwidth}{!}{\rotatebox{0}{\includegraphics{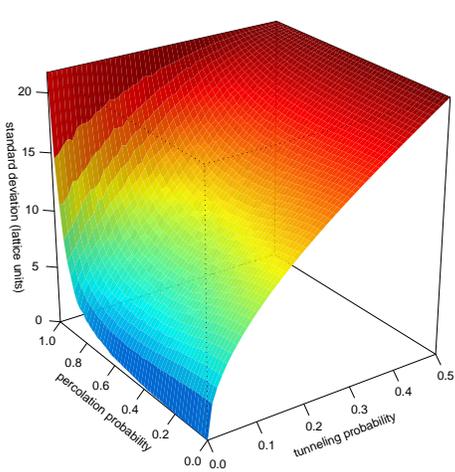}}}
    \end{center}
  {(a) $\langle x_{\text{rms}} \rangle$ for static gaps.}
  \end{minipage}
\hfill
  \begin{minipage}{0.48\columnwidth}
    \begin{center}
        \resizebox{0.75\columnwidth}{!}{\rotatebox{0}{\includegraphics{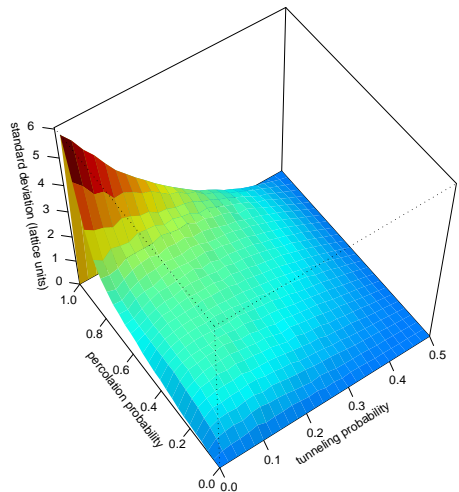}}}
    \end{center}
  {(b) $\sigma(x_{\text{rms}})$ for static gaps.}
  \end{minipage}

\vspace{2em}

  \begin{minipage}{0.48\columnwidth}
    \begin{center}
        \resizebox{0.75\columnwidth}{!}{\rotatebox{0}{\includegraphics{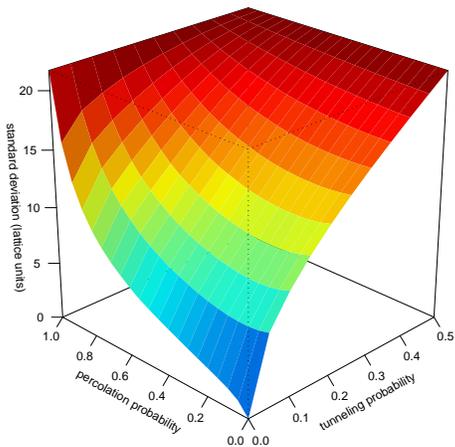}}}
    \end{center}
  {(c) $\langle x_{\text{rms}} \rangle$ for dynamic gaps.}
  \end{minipage}
\hfill
  \begin{minipage}{0.48\columnwidth}
    \begin{center}
        \resizebox{0.75\columnwidth}{!}{\rotatebox{0}{\includegraphics{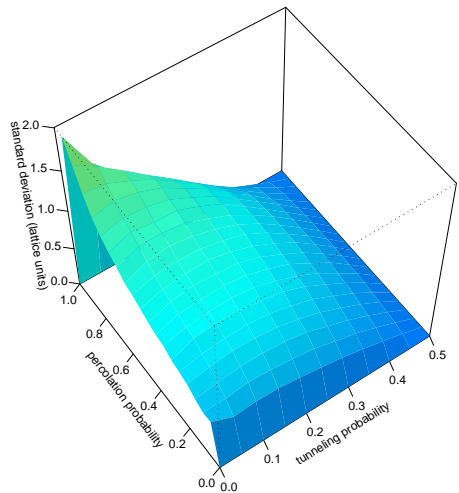}}}
    \end{center}
  {(d) $\sigma(x_{\text{rms}})$ for dynamic gaps.}
  \end{minipage}

\end{minipage}
\caption{Plots of $\langle x_{\text{rms}} \rangle$ (left) and $\sigma(x_{\text{rms}})$ (right) against
	$0<\eta<0.5$ and $0<p<1.0$ for a 1D quantum walk,
	with static (top) and dynamic (bottom),
	run for 40 steps, and averaged over
	1000 random gap arrangements.
	Note scale mismatches in vertical axes.}
\label{fig:1DSDS}
\end{figure}

For a random arrangement of missing edges that stays constant
throughout the walk, $\langle x_{\text{rms}} \rangle$
and $\sigma(x_{\text{rms}})$
are shown in figures \ref{fig:1DSDS}(a) 
and \ref{fig:1DSDS}(b). 
The $\langle x_{\text{rms}} \rangle$ in figure \ref{fig:1DSDS}(a) 
is as one would expect, with the gaps having a big impact
on the spreading for low
tunneling rates, while even a small amount of tunneling
allows the spreading rate to steadily recover.
The $\sigma(x_{\text{rms}})$ in figure \ref{fig:1DSDS}(b)
shows that this behaviour is reliable (low variation)
except where both the number of gaps and the tunneling rate are low.

For dynamic gaps, the equivalent pair of plots are shown in figures
\ref{fig:1DSDS}(c) 
and \ref{fig:1DSDS}(d). 
The introduction of dynamic gaps can be seen to have a steep but
less drastic effect than static gaps by comparing 
figure \ref{fig:1DSDS}(a) 
with figure \ref{fig:1DSDS}(c). 
The $\sigma(x_{\text{rms}})$ is also lower, indicating the
behaviour is less variable between different lattices with the 
same percolation probability $p$.
This is due to the dynamic gaps providing an averaging  effect
during the course of the walk.

The main message from these results for quantum walks on a line
with edges missing is that a random pattern of missing edges
leads to decoherence in the quantum walk, reducing the spreading
to the classical random walk scaling proportional to $\sqrt{t}$, but
with varying prefactors.  However, the crossover to classical scaling 
happens quite slowly, after a crossover time proportional
to $1/(1-p)$ as found by \citeauthor{romanelli03a}
\cite{romanelli03a} for the case of dynamic gaps.
This means the decoherence effects are minimal for smaller systems
allowing the full quantum speed up to be exploited, while larger
systems will only benefit from a prefactor enhancing the classical
scaling.


\section{Walk on a two-dimensional grid}
\label{sec:percgrid}

A two-dimensional lattice allows a wider range of behaviour to be studied,
without resorting to quantum tunneling.  Since there is more than one
path between two points on the lattice,
the walker can take a (probably longer) path around the gaps when they
are at a low density.
Each vertex in a 2D lattice has four edges connected to it, so
the quantum walker now needs a four-dimensional coin.  We label the four
directions $\ket{L}$, $\ket{R}$ corresponding to $\ket{-1}$ and $\ket{+1}$
on the $x$ axis and $\ket{D}$, $\ket{U}$ for $\ket{-1}$ and $\ket{+1}$
on the $y$ axis.
The operator for ``tossing'' the coin can now be any $4\times 4$ unitary.
The most common choice is based on Grover's diffusion operator,
\begin{equation}
C_4^{(Grov)} = \frac{1}{2}\left( \begin{matrix} -1 & 1 & 1 & 1 \\ 1 & -1 & 1 & 1 \\ 1
 & 1 & -1 & 1 \\ 1 & 1 & 1 & -1 \end{matrix} \right ).
\label{eq:Grover}
\end{equation}
The general form of Grover's diffusion operator for a $d$-edged vertex
is $2/d - \identity_d$, it happens to be unbiased only for $d=4$.
It is chosen because of its symmetry, the incoming edge receives
a sign flip, but the other three edges are treated equally.  There are
many possible unbiased coin operators based on generalised Hadamard
matrices \cite{tadej06a}, but these will have other phases that
treat the outgoing directions differently.

To complement this choice of coin operator, the shift operator $\mathbf{S}_4$
must also invert the coin states,
\begin{eqnarray}
&&\mathbf{S}_4\ket{L,x,y} = \ket{R,x-1,y}\nonumber\\
&&\mathbf{S}_4\ket{R,x,y} = \ket{L,x+1,y}\nonumber\\
&&\mathbf{S}_4\ket{D,x,y} = \ket{U,x,y-1}\nonumber\\
&&\mathbf{S}_4\ket{U,x,y} = \ket{D,x,y+1}.
\end{eqnarray}
This ensures that the coin operator is then applied to a coin state
that labels the edge from which the walker arrived,
which is the one that receives the minus sign from the Grover
coin operator.  For example,
\begin{equation}
C_4^{(Grov)}\ket{L,x,y} = \frac{1}{2}\left(-\ket{L,x,y} + \ket{R,x,y}
	+ \ket{D,x,y} + \ket{U,x,y} \right).
\end{equation}

\begin{figure}[tbh!]
  \begin{minipage}{0.48\columnwidth}
    \begin{center}
        \resizebox{0.75\columnwidth}{!}{\rotatebox{0}{\includegraphics{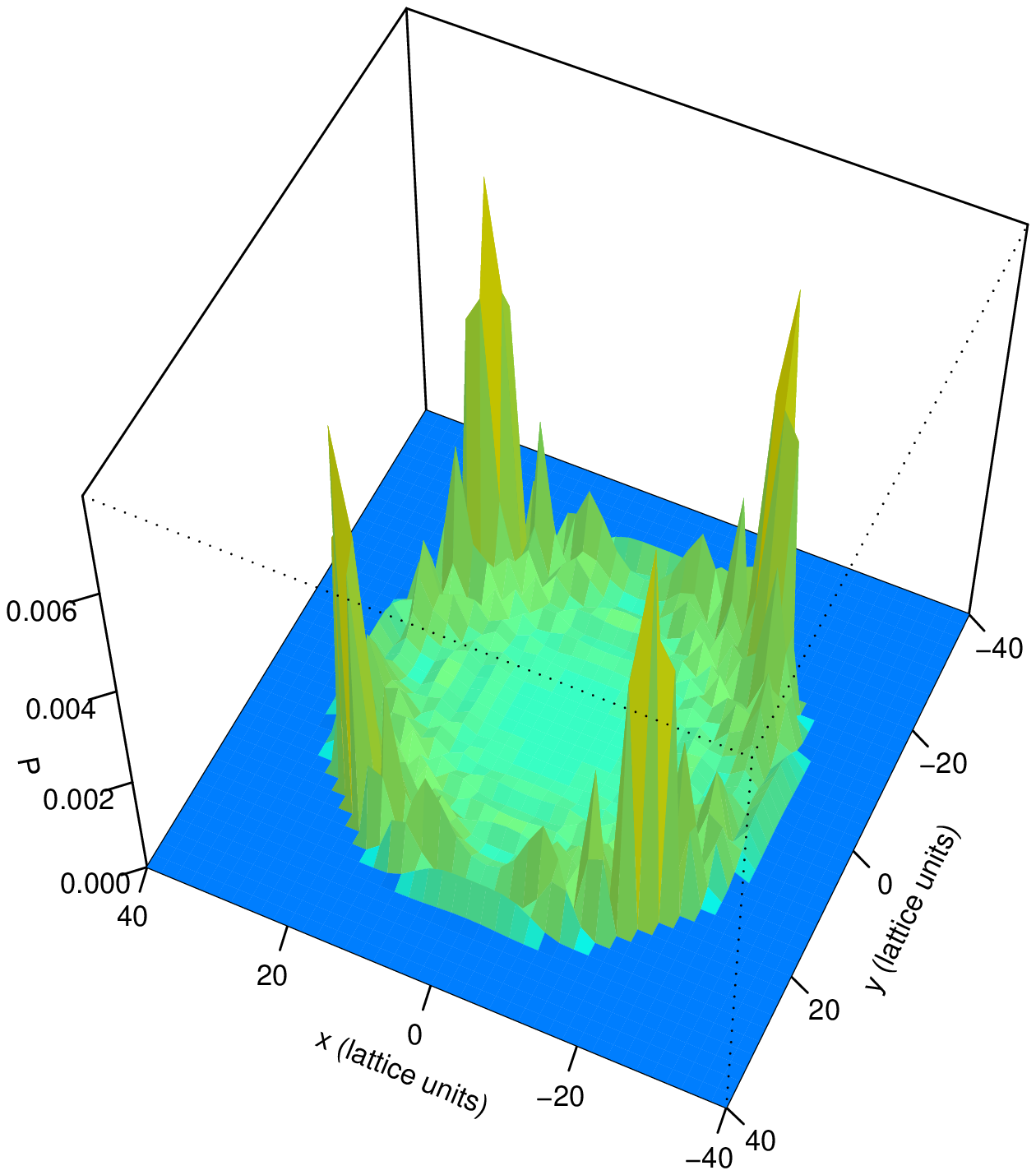}}}
    \end{center}
  \end{minipage}
  \hfill
  \begin{minipage}{0.48\columnwidth}
    \begin{center}
        \resizebox{0.75\columnwidth}{!}{\rotatebox{0}{\includegraphics{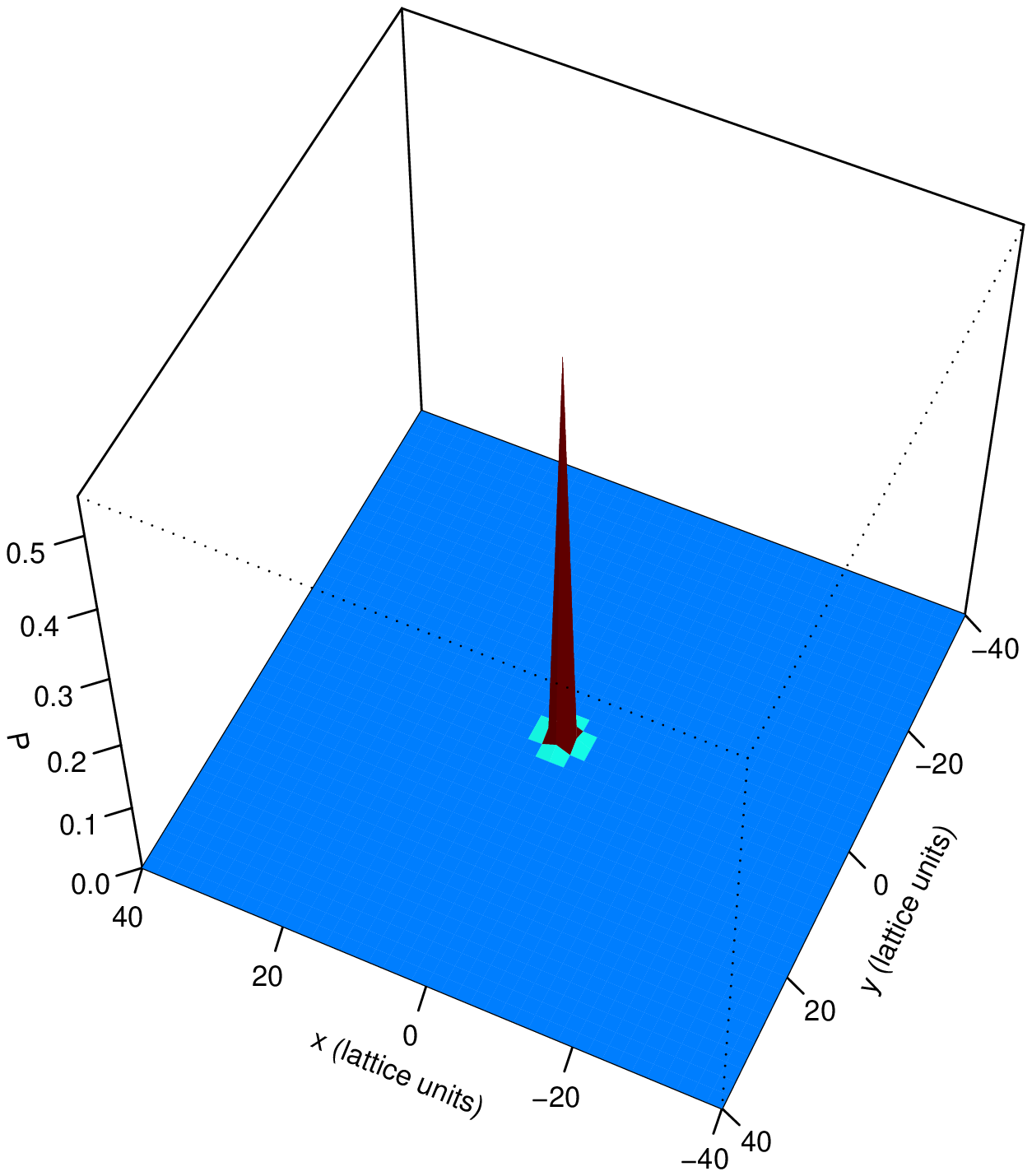}}}
    \end{center}
  \end{minipage}
  \caption{Probability distribution for quantum walk on 2D lattice run for
	40 steps with initial states
	$0.5(\ket{L,0}+\ket{R,0}+\ket{U,0}+\ket{D,0}$ (left)
	and $0.5(-\ket{L,0}-\ket{R,0}+\ket{U,0}+\ket{D,0}$ (right).
	Only even positions are shown since odd positions are unoccupied.
	Note scale mismatch in vertical axes.}
  \label{fig:T40ProbI3}
\end{figure}

First we recall the behaviour of a quantum walk on a lattice with all
edges present.  This is illustrated in figure 
\ref{fig:T40ProbI3}. 
For most choices of coin and initial state,
the quantum walk tends to stay around the origin.  For a well-behaved coin 
(with a good degree of symmetry), there is one precisely chosen 
state for which it spreads out in a ring of variable height that resembles
a circular version of the walk on the line \cite{tregenna03a}.
If we start the walker off from the origin ($x=0$, $y=0$),
we can use the average distance of the walker from the origin $\bar{r}$,
the average of $r=\sqrt{x^2+y^2}$, as a measure of the spreading.
This is playing the same role as $x_{\text{rms}}$
for the walk on the line, although $\bar{r}$ is more
directly comparable with the average of $|x|$,
since the square root is taken before the averaging.
In figure \ref{fig:T40ProbI3}, the cases with the two extreme values 
of $\bar{r}$ for the Grover coin
operator are shown, on the left $\bar{r}$ after 40 steps is 33.1,
while on the right it is only 7.2.  Nonetheless, this still represents
a quantum spreading rate, despite the apparent dominance of the
central peak.  The classical spreading of $\sqrt{t}$ is only 6.3
for 40 steps.
Other initial states show a mixture
of the ring and central peak, with corresponding intermediate
spreading rates.  However, the central peak is present and
dominates over the ring for all except the symmetric starting state.
This is in sharp contrast with 
the walk on the line where the spreading rate 
is the same for all initial coin states,
only the symmetry of the distribution changes.
Analytic treatment of the quantum walk in two dimensions has
been provided by \citeauthor{grimmett03a} \cite{grimmett03a}
and \citeauthor{gottlieb04a} \cite{gottlieb04a}.

\subsection{Two-dimensional percolation lattices}

Applying a quantum walk to a two-dimensional percolation lattice
is a straightforward extension of the techniques already 
described for varying the dynamics of the walk on the line.
For both bond and site percolation, the resulting lattice
will in general contains vertices of degree 4, 3, 2, 1 and 0.
Each needs an appropriate coin operator.
We constructed these in the most symmetric possible way,
basing the lower dimensional coins on the lower dimensional
Grover operators padded with (minus) the identity for the 
directions where the edges are missing.  For example,
if the $\ket{U}$ edge is missing, the coin operator
becomes
\begin{equation}
C_4^{(U)} = \frac{1}{3}\left( \begin{matrix} -1 & 2 & 2 & 0 \\ 2 & -1 & 2 & 0 \\
2 & 2 & -1 & 0 \\ 0 & 0 & 0 & -3 \end{matrix} \right ),
\end{equation}
while if both the $\ket{U}$ and $\ket{L}$ edges are missing,
\begin{equation}
C_4^{(LU)} = \left( \begin{matrix} -1 & 0 & 0 & 0 \\ 0 & 0 & 1 & 0 \\
0 & 1 & 0 & 0 \\ 0 & 0 & 0 & -1 \end{matrix} \right ),
\end{equation}
Provided the initial state is correctly chosen, so the subspace of the
missing edges is never populated, it doesn't matter what appears in
those rows and columns of the coin operators.  However, if one wishes to 
interpolate between missing edges and the full lattice, in a 
generalisation of the tunnelling studied on the line in the previous
section, the sign of the entries on the diagonal must be the same
throughout the coin operator, as shown above.  For consistency,
the coin operator for all but one edge missing 
\begin{equation}
C_4^{(id)} = \left( \begin{matrix} -1 & 0 & 0 & 0 \\ 0 & -1 & 0 & 0 \\
0 & 0 & -1 & 0 \\ 0 & 0 & 0 & -1 \end{matrix} \right ).
\end{equation}
applies a minus sign to the amplitude before it returns along the edge.

Since the initial coin state has a significant effect
on the spreading rate of the walk, 
we therefore tested several different initial coin states,
to see how this affected
the behaviour, in combination with the missing edges in
the percolation lattices.
We tested the initial states $\ket{\psi_{\text{max}}}$ and
$\ket{\psi_{\text{min}}}$ that give the maximum and minimum
spreading rates as shown in figure \ref{fig:T40ProbI3},
\begin{equation}
\ket{\psi_{\text{max}}} = 0.5(\ket{L,0}+\ket{R,0}+\ket{U,0}+\ket{D,0},
\label{eq:psimax}
\end{equation}
and 
\begin{equation}
\ket{\psi_{\text{min}}} = 0.5(-\ket{L,0}-\ket{R,0}+\ket{U,0}+\ket{D,0}.
\label{eq:psimin}
\end{equation}
These differ only in the phases between the different coin directions.
We thus also tested random phases of $\pm1$ between the different coin
directions,
\begin{equation}
\ket{\psi_{\text{ran}}} = 0.5(\pm\ket{L,0}\pm\ket{R,0}\pm\ket{U,0}\pm\ket{D,0}.
\label{eq:psiran}
\end{equation}
This is a similar strategy to that used by \citeauthor{tregenna03a}
\cite{tregenna03a} when obtaining the maximum and minimum spreading rates.
Restricting the variation to phases avoids skewing the walk, which would 
systematically increase the average distance traveled, making it harder
to extract the generic average behaviour from the results.
In all cases, the walker started at the origin.  If there happened
to be missing edges, those coin states were deleted from the
initial state, and the remainder renormalized.  If all the edges
happened to be missing around the origin, a spreading of $\bar{r}=0$
was recorded, since no walk was possible.

For each initial state in equations (\ref{eq:psimax}) to (\ref{eq:psiran}),
we performed a numerical simulation of a quantum walk on both site and
bond percolation lattices, averaging over 5000 randomly generated
lattices of each type.  We were able to run up to 140 time steps on a
3GHz processor in reasonable time (a week per 5000 sample lattices
per initial state).  
With $(2\times140+1)^2= 78961$ lattice sites, this is equivalent in memory
requirements to nearly 40000 steps of a walk on a line.
Arguably, 5000 sample lattices is not a sufficiently large sample for such
large lattices.  We investigated the sensitivity to sample size by 
comparing averages for smaller lattices (40 to 80 time steps) for 1000, 5000 
and 20000 samples.  The larger samples produced smoother averages as
expected, but the smaller samples gave good enough results to draw 
conclusions from.

\begin{figure}[tbh!]
  \begin{minipage}{1.0\columnwidth}
    \begin{center}
        \resizebox{0.65\columnwidth}{!}{\rotatebox{0}{\includegraphics{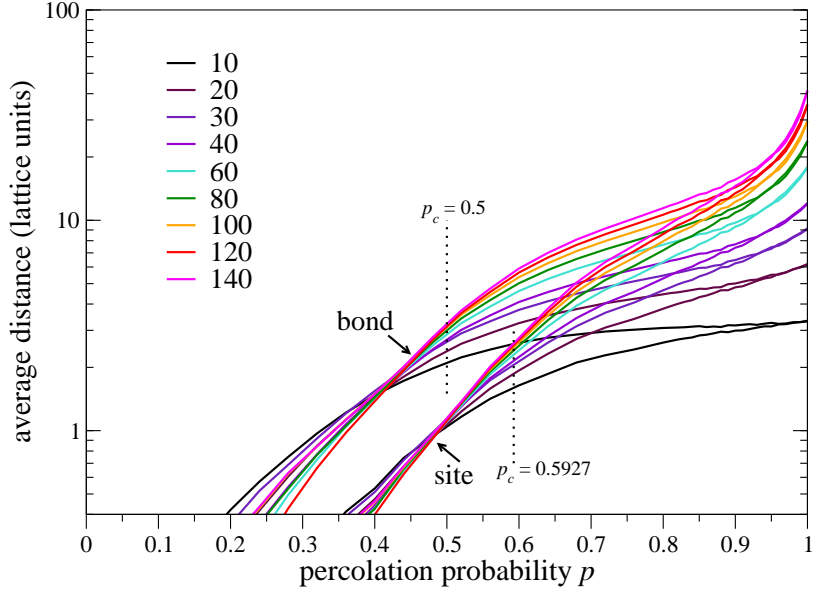}}}
    \end{center}
  \end{minipage}
  \caption{Comparison of the average distance $\langle\bar{r}\rangle$
	from the starting point for bond (upper set) and site (lower set)
	percolation for the number of steps from 10 to 140 as shown in the key,
	as a function of the percolation probability $p$,
	with initial states using random phases, equation (\ref{eq:psiran}).}
  \label{fig:bond+site-rp}
\end{figure}

\begin{figure}[tbh!]
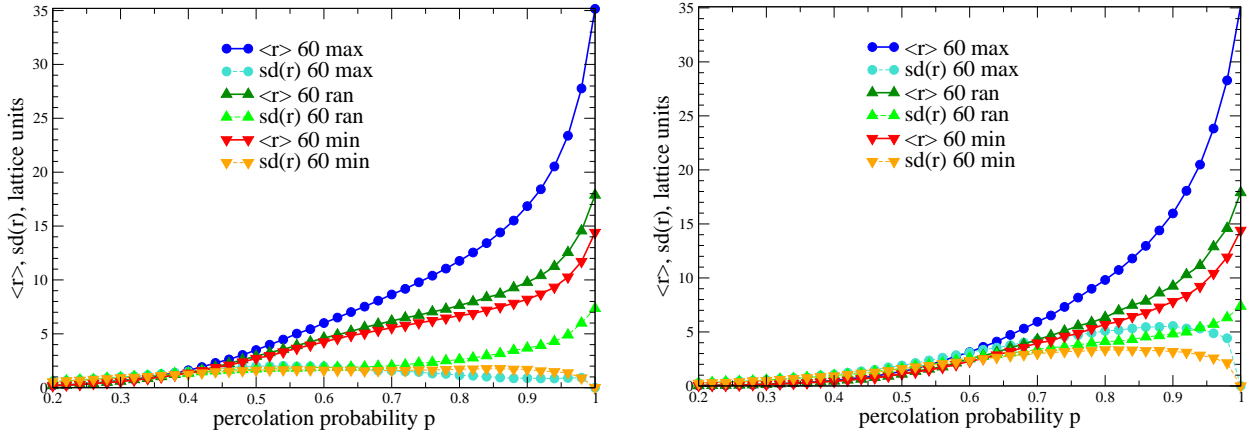

  \begin{minipage}{0.48\columnwidth}
    \begin{center}
        \resizebox{1.00\columnwidth}{!}{\rotatebox{0}{\includegraphics{eps/bond60.eps}}}
    \end{center}
  \end{minipage}
  \hfill
  \begin{minipage}{0.48\columnwidth}
    \begin{center}
        \resizebox{1.00\columnwidth}{!}{\rotatebox{0}{\includegraphics{eps/site60.eps}}}
    \end{center}
  \end{minipage}
  \caption{Comparison of the mean distance $\langle\bar{r}\rangle$
	from the starting point for bond (left) and site (right) percolation
	over 60 steps, averages over 5000 random percolation lattices,
	as a function of the percolation probability $p$,
	with initial states giving the maximum and minimum distance
	in the full lattice case, plus averaging over random phases in
	the initial state. The standard deviation $\sigma(\bar{r})$
	over the random percolation lattices is also shown.}
  \label{fig:perc60}
\end{figure}

\begin{figure}[tbh!]
 \begin{minipage}{1.00\columnwidth}
  \begin{minipage}{0.48\columnwidth}
    \begin{center}
        \resizebox{0.85\columnwidth}{!}{\rotatebox{0}{\includegraphics{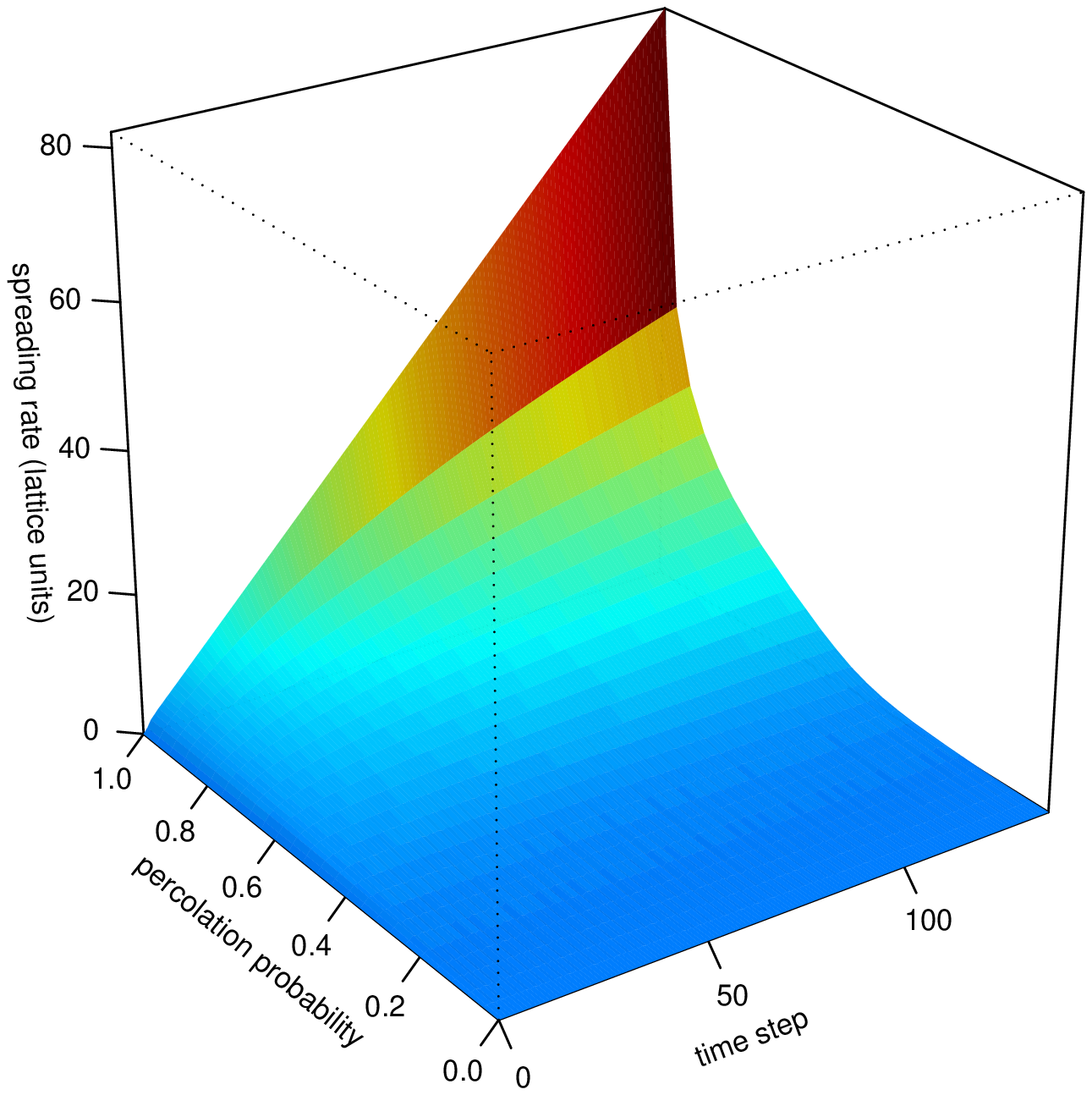}}}
    \end{center}
  \end{minipage}
  \hfill
  \begin{minipage}{0.48\columnwidth}
    \begin{center}
        \resizebox{0.85\columnwidth}{!}{\rotatebox{0}{\includegraphics{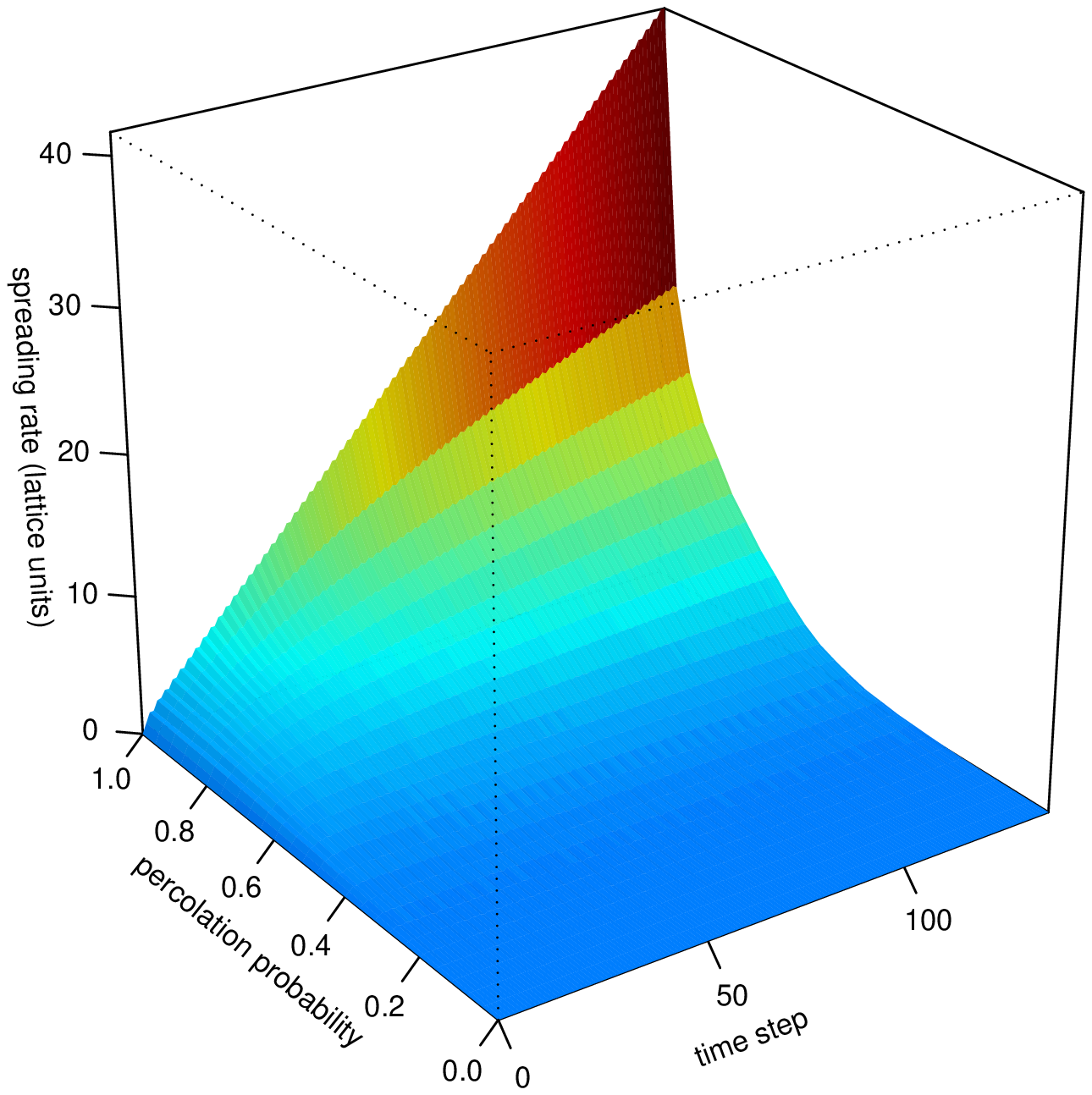}}}
    \end{center}
  \end{minipage}
  \caption{Variation of spreading rate $\langle\bar{r}\rangle$
	with percolation probability
	up to 140 steps for bond (left) with max spread initial state,
	and site (right) percolation on 2D Cartesian lattice with random
	initial state. Note difference in vertical scales.}
  \label{fig:siterp140}
 \end{minipage}
\end{figure}

Figures \ref{fig:bond+site-rp}, \ref{fig:perc60} and \ref{fig:siterp140}
present the results of these simulations in three complementary ways.
In figure \ref{fig:bond+site-rp}, we show the results for quantum walks
with random phases in their initial states.
The average distance from the starting position at the origin
for a single quantum walk, $\bar{r}$ where $r=\sqrt{x^2+y^2}$,
is given by 
\begin{equation}
\bar{r} = \sum_{x,y} \sqrt{x^2+y^2}|\psi(x,y)|^2.
\label{eq:avr}
\end{equation}
Here $|\psi(x,y)|^2$ is the probability of finding the particle
at the site $(x,y)$, i.e., summed over all coin states,
and the sum over $x,y$ is over all lattice sites that could
be reached by the walker, i.e., $-t \le x,y \le t$.
Clearly, $\bar{r}$ can vary, depending on the exact arrangement of
missing sites or edges in the percolation lattice.  We have therefore
further averaged over 5000 random instances of bond or
site percolation lattices, and plotted the resulting 
$\langle\bar{r}\rangle$ as a function of the percolation
probability $p$, for different lengths of quantum walk from 10 to 140 steps.
Values of $\langle\bar{r}\rangle$ below one that occur for small
$p$ are not interesting: the low density of sites or edges in the lattice
means that most of the time the quantum walk is unable to move from its
starting state.
While there is clearly some noise in the data, particularly for site
percolation, the trends are clear.  At around $p=0.4$ for bond
($p=0.5$ for site) percolation, $\langle\bar{r}\rangle$ starts
to rise.  This is still below the critical probability $p_c=0.5$
($p_c=0.59\dots$) and the value of $\langle\bar{r}\rangle$ is
not sensitive to the number of time steps because in general it is
still confined to a small local region and cannot see the full size of
the lattice.
We then observe the lines for different time steps separating out
below the main bundle in succession, smallest first, and flattening off 
towards the corresponding value for the complete lattice at $p=1$.
This is a nice illustration of finite size effects.  The size of the
localized clusters grows as $p$ approaches $p_c$ from below.  As
the cluster size becomes comparable with the size of the walk (10, 20
time steps, etc.), the limitation of $\langle\bar{r}\rangle$
due to cluster size is replaced by limitation due to the number of steps,
and the walk is no longer sensitive to increasing cluster size.
Walks of about 100 steps or longer are still close to synchronized at
$p=p_c$, indicating that quantum walks of this size should
provide good estimates of critical parameters.

In the region where $p_c<p<1$, the rate of growth of
$\langle\bar{r}\rangle$ slows then steepens again.
For $p_c\gtrsim 0.95$, the walk no longer distinguishes between
bond and site percolation, $\langle\bar{r}\rangle$
is the same for both.  This is somewhat surprising, because removing
a site from the lattice removes four neighbouring edges, compared 
to removing a single edge for bond percolation.  It suggests that
in the regime where missing sites or edges can be expected to be
isolated from each other, they both behave as lattice defects,
with the broken lattice symmetry more important that the detailed
structure of the defect.

To understand this behaviour in more detail, we turn to 
figure \ref{fig:perc60}, where walks of 60 time steps are
compared for all three initial states.
Also plotted is $\sigma(\bar{r})$, the standard deviation of
$\bar{r}$, to quantify how variable the spreading is
between different percolation lattices.
Note first that the average spreading $\langle\bar{r}\rangle$
for $p=1$ ranges from around 14 for $\ket{\psi_{\text{min}}}$ to 35 for 
$\ket{\psi_{\text{max}}}$.  The value for $\ket{\psi_{\text{ran}}}$
is around 18, half that for $\ket{\psi_{\text{max}}}$.  However, 
 $\sigma(\bar{r})$ for $\ket{\psi_{\text{ran}}}$ is around seven, 
indicating that the average value conceals significant variation over the
random phases.  This is as expected, since the randomly chosen phases 
should include both $\ket{\psi_{\text{min}}}$ and $\ket{\psi_{\text{max}}}$.
The position of $\ket{\psi_{\text{ran}}}$ close to $\ket{\psi_{\text{min}}}$
indicates the weight of the distribution lies nearer the minimum
spreading rather than the maximum.  All this is already known from
previous studies, of course, but it is useful to recall the details
before considering the behaviour for $p<1$.

Looking at the standard deviations, it can be seen that the variation
for $\ket{\psi_{\text{ran}}}$ is larger for site percolation, typically
around five compared with around two for bond percolation for $p_c<p<0.95$.
Thus, although the average behaviour is similar, missing sites 
produce more extreme cases (high and low) for the same number
of missing bonds in bond percolation.
This suggests further work to elucidate the source of this
difference would be fruitful.

The variation of $\langle\bar{r}\rangle$ over time with 
$p$ can be plotted in 3D, see figure
\ref{fig:siterp140} for two examples.
This shows how similar the behaviour is for both bond and site percolation,
with any choice of initial state, but it does not bring out the
most interesting features to do with how the scaling varies with the 
percolation probability.  The necessary scaling analysis is
described in the following subsection.

\subsection{Scaling analysis}

\begin{figure}[tbh!]
  \begin{minipage}{0.98\columnwidth}
    \begin{center}
        \resizebox{0.75\columnwidth}{!}{\rotatebox{0}{\includegraphics{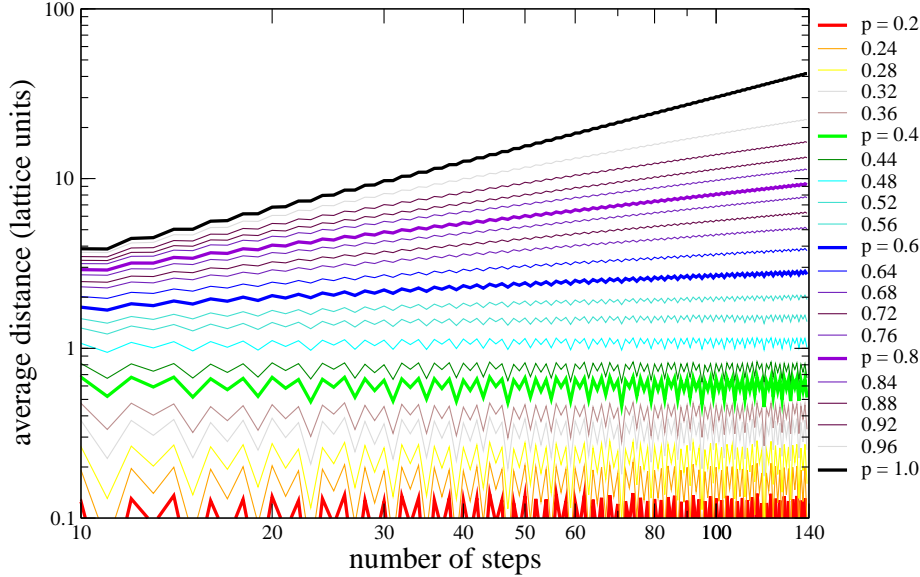}}}
    \end{center}
  \caption{Variation of spreading rate $\langle\bar{r}\rangle$
	with percolation probability
	up to 140 steps for site percolation on 2D Cartesian lattice
	with random initial state.  A log-log plot so slope gives $\alpha$
	where $\langle\bar{r}\rangle \sim t^{\alpha}$.}
  \label{fig:siterp140ll}
  \end{minipage}
\end{figure}

\begin{figure}[tbh!]
  \begin{minipage}{0.98\columnwidth}
    \begin{center}
        \resizebox{0.75\columnwidth}{!}{\rotatebox{0}{\includegraphics{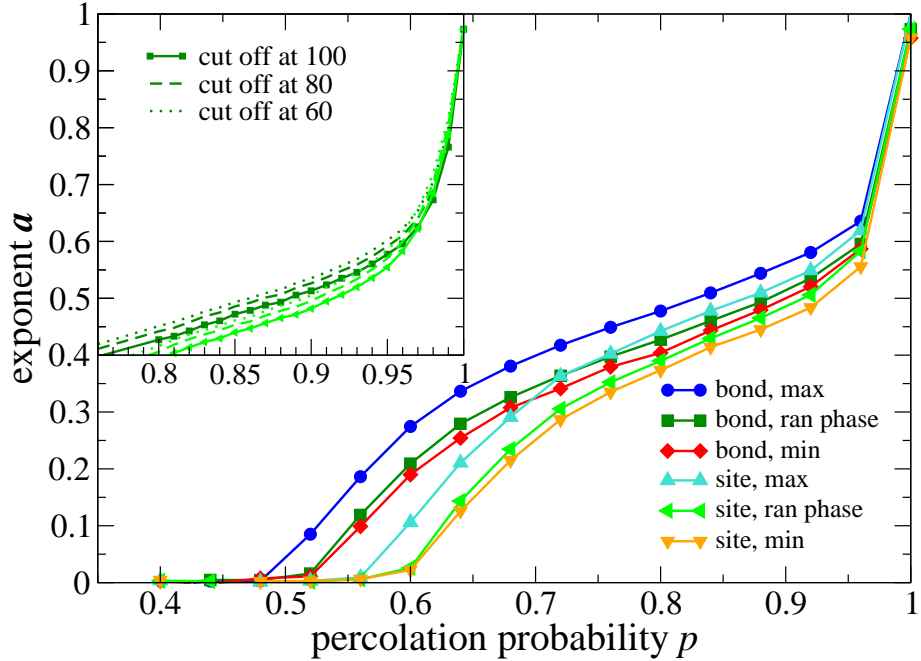}}}
    \end{center}
  \caption{Fractional scaling exponent for 2D percolation lattices
	derived from from data for $t=100$ to $140$ steps.
	Initial states (bond, site) for maximum (blue, cyan),
	minimum (red, orange) spreading and random phases (green, lime).
	Inset shows greater detail for $0.8<p<1.0$ for random phases,
	and results of using lower cut off of 80 (dashed) or 60 (dotted) steps
	to determine exponent.}
  \label{fig:slopefits}
  \end{minipage}
\end{figure}

A log-log plot of the same data in figure \ref{fig:siterp140} (right)
is shown in 2D in figure \ref{fig:siterp140ll}.  
Apart from a regular zigzag over a cycle of four steps, due to the lattice
and coin symmetries, the line for a fixed value of $p$ is very close to
a straight line, with slope varying from zero (small $p$) to one ($p=1$).
This suggests the data will be well fit by $\langle\bar{r}\rangle
\propto t^{\alpha}$ with $0\le\alpha\le 1$.
Mindful of the finite size effects discussed in relation to figure
\ref{fig:bond+site-rp}, fits to determine $\alpha$ were done using
only the data for time steps 100 to 140.  The resulting values for 
the exponent $\alpha$ are shown in figure \ref{fig:slopefits}.
The values of $\alpha$ agree fairly well with the
expectation that the first rise above zero should be seen around
the critical point $p_c=0.5$ (bond) and $p_c=0.59\dots$ (site).
The initial state that produces maximum spreading has a noticeably 
higher exponent, around 0.05 larger than for random phases or the 
minimum spreading state, from the critical point through to near $p=1$,
implying that the advantage of the optimal initial state is not just
a prefactor on the same scaling rate.
All initial states follow the same overall pattern of a steady rise 
from the critical point, a flattening off towards $\alpha=0.5$,
the classical exponent for the perfect lattice, in the region of 
$p=0.9$, followed by a steep rise to the quantum spreading rate of
$\alpha=1$.

To test the influence of finite size effects on these results,
the same fits were done using more of the data, with $t=80$
and $t=60$ as lower cut off points in place of $t=100$.
The slightly higher exponents these fits produce are shown as an inset
in figure \ref{fig:slopefits}, along with more detailed data,
for the case with random phase initial states.
These show that the trend towards lower values of $\alpha$ is
slow, and hence, based on these simulations,
we cannot make a strong prediction for what the asymptotic variation
of the exponent with percolation rate is going to be.
Nonetheless, we can confidently say that for systems of limited size,
there is a significant window for $p\gtrsim0.9$
in which the quantum advantage exists
in the form of faster spreading than a classical random walk.

\section{Summary and discussion}
\label{sec:summary}

Even for the simple case of a coined quantum walk on the line
we find contrasting effects brought about by the interplay between
missing edges and quantum tunnelling, in the presence of
ordered or disordered arrangements of the missing edges.
Ultimately, for long times, the disorder of random missing
edges causes the spreading to reduce to the classical
$\sqrt{t}$ rate.  Yet for shorter walks, even up to a thousand
steps, the spreading is still largely dominated by quantum effects.
And for ordered arrangements of missing edges, the quantum
spreading rate is maintained, though with varying prefactors.
These results are consistent with those of a number of other
analytical and numerical studies, in particular
\citeauthor{romanelli03a} \cite{romanelli03a} who first
studied randomly changing gaps on the line, and 
\citeauthor{linden09a} \cite{linden09a}, who provide detailed
analytic treatment of periodically varying coins.
The connections between disorder and decoherence were discussed
in detail by \citeauthor{romanelli03a}, highlighting the role
of randomness from whatever source in degrading the quantum coherence.
The scaling of quantum walks on the line with decoherence or
other sources of disorder are in general either linear spreading or
$\sqrt{t}$ spreading, with different prefactors depending on the
details of the dynamics.  Intermediate scaling in one-dimensional
walks is not obtained except transiently as linear crosses over
to a long time limit of $\sqrt{t}$.

In contrast to the 1D cases, on 2D percolation lattices
we find the quantum walks show fractional scaling of the
spreading, i.e., $\bar{r}\propto t^{\alpha}$ for $p_c < p < 1$,
with $0\le\alpha\le 1$.
Classical random walks on percolation lattices also exhibit
fractional scaling behaviour, with $0\le\alpha\le0.5$.
One interesting question is whether this behaviour seen for
finite-sized walks extends to the long time limit.
Although our simulations seem to have reached a size where
finite size effects are small, the slow convergence, coupled with
the lessons from one dimensional quantum walks where 
decoherence dominated only after thousands of steps, means
we do not think we can confidently predict the large $t$ behaviour.
For example, a possibility consistent with our results is that
the steep rise in the region of $0.95\le p\le1.0$
will becomes a ``step'' function at $p=1$ as
$t\rightarrow\infty$, and that the portion below $\alpha=0.5$
will follow the scaling for classical random walks.
The intuition for this is that the randomness in the percolation lattice
would thus again act as decoherence in the large $t$ limit.
Whether this is correct depends on whether a very small density of defects is
enough to disrupt the entire walk in the whole two-dimensional lattice
in the long time limit.  Studies of absorbing boundaries and trapping
are relevant to this question,
in one dimension the presence of traps does not
inevitably catch a quantum walker \cite{bach02a},
while a classical walker will
eventually fall into a single trap in both one and two dimensions.
This will be an interesting question to answer because
trapping behaviour is particularly important for
models of exciton transport, as discussed by
\citeauthor{muelken07a} \cite{muelken07a},
and trapping can also be
exploited to guide the exciton to the desired location.
\citeauthor{feldman03a} \cite{feldman03a} and
\citeauthor{hillery03a} \cite{hillery03a} provide a scattering matrix
method for treating finite graphs analytically, and determining
solutions that traverse the graph or are trapped within it.
This may provide a route to understanding how structures in
the percolation lattice contribute to trapping the quantum walker.

\begin{figure}[tbh!]
  \begin{minipage}{0.98\columnwidth}
    \begin{center}
        \resizebox{0.75\columnwidth}{!}{\rotatebox{0}{\includegraphics{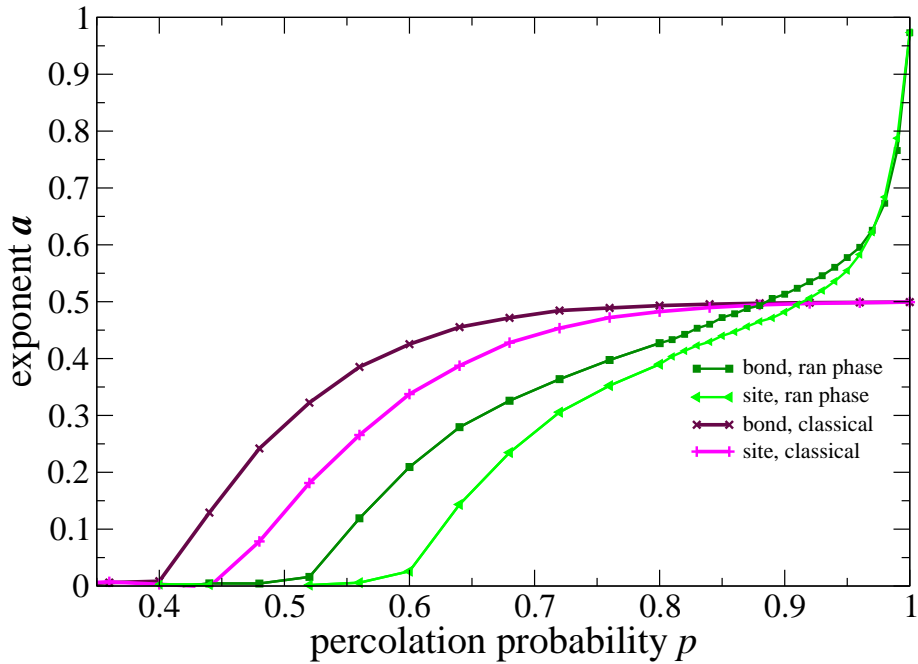}}}
    \end{center}
  \caption{Fractional scaling exponent for 2D percolation lattices
	derived from from data for $t=100$ to $140$ steps.
	Initial states (bond, site) for random phases (green, lime)
	are shown (same data as in figure \ref{fig:slopefits}).
	Classical random walk analysed in the same manner is
	shown for comparison (maroon, magenta), which displays
	obvious finite size effects.}
  \label{fig:slopefitclass}
  \end{minipage}
\end{figure}
Aside from the question of trapping behaviour, 
we offer some further intriguing observations that 
suggest directions for future study.  Figure \ref{fig:slopefitclass}
shows the data for random initial states from figure \ref{fig:slopefits}
plotted along with numerical results for running a classical
random walk on percolation lattices up to 140 steps, and processed to
extract the scaling in exactly the same way.
The first and most obvious problem is that 140 steps of a classical random
walk still suffers from large finite size effects!  This is easily 
understood, $\bar{r}$ for a classical random walk on the fully
connected lattice is $\sqrt{t}$, which is about 12 for $t=140$.  This
is more comparable with $t=40$ for a quantum walk (with random phases
in the initial state), as can be obtained from figure \ref{fig:bond+site-rp},
reading off the values for $p=1$.
Since $\bar{r}$ gives an indication of 
the size of the largest structures that the quantum or random walk will be
sensitive to, if the missing sites or edges are on average
further apart than that, the walker will hardly notice any before
finishing the walk and reporting its position.
Working in the other direction, a quantum walk of 100 steps (with random
phase initial state) has $\bar{r}=30$.  If we want
a classical random walk with $\bar{r}=30$,
will need to run it for $30\times 30 = 900$ steps!  The advantage of the
linear scaling with $t$ of the quantum walks is suddenly very apparent.

As already noted, we appear to have achieved a reasonable
convergence of finite size effects on percolation lattices
by running a classical simulation of a quantum walk for 100 or more steps.
These simulations required around a week of processor
time on a fast workstation, a fairly modest amount of computational
resources.  This suggests that using such simulations to obtain 
numerical values for percolation lattice parameters is worth
investigating.  In other words, this could provide a route to 
improved classical simulation methods.
There are other examples of quantum algorithms
providing better classical methods, \citeauthor{love06a}
\cite{love06a} discuss a number of examples in detail.

However, the main reason for running the classical random walk was to 
compare with the quantum walk and further understand our results.
Much larger classical simulations were not possible with our
available computational resources,
we would have needed an order of magnitude more computational
power, or to implement more efficient simulation techniques.
Comparisons using the results in figure \ref{fig:slopefitclass}
can only be suggestive, nonetheless, the suggestions are
intriguing.
If we assume that the shape of the classical curves in figure
\ref{fig:slopefitclass} won't change much, so
they will simply shift to the right as finite size effects reduce,
then we can compare the shape with the curves for quantum walks.
For site percolation, the classical and quantum curves are
the same shape up to about $p=0.9$ on the quantum curve.
The corresponding curves for bond percolation are not, however, the
quantum curve lies significantly below the classical curve for the region
$0.55<p<0.85$.  This adds further evidence for basic
differences in quantum walk behaviour on bond and site percolation
lattices that the differences in standard deviation noted in 
figure \ref{fig:perc60} imply.
In particular, that quantum walks are slower than classical in some regions,
while faster than classical in others.  Such behaviour has been
noted in other contexts, see for example \citeauthor{agliari08a}
\cite{agliari08a}.

Nonetheless, the main message we want to emphasize from this study is that
for modest system sizes of a few hundred sites, and defect densities
below 10\%, faster-than-classical fractional scaling is very much the
dominant feature.  This is an interesting regime with many applications
in both bio- and nano-scale materials, where quantum effects are
important to fully understand their properties and behaviour.

\section*{Acknowledgments}

We thank Jiajun Tan for assistance in preparing the references for this paper.
VK is funded by a Royal Society University Research Fellowship.  
PK was funded by a Nuffield Undergraduate Summer Bursary.
JB was funded by a Royal Society Summer Bursary.
GLMC was funded by the University of Leeds School of Physics and Astronomy 
Summer Bursary scheme.



\renewcommand{\baselinestretch}{1.0}\small\normalsize
\bibliography{../bibs/qrw,perc,../bibs/misc,../bibs/qbio,../bibs/qit}



\end{document}